    \documentclass{aastex}
   \usepackage{emulateapj5}
   
\begin{document}
   
   \title{Evolution of the  Correlation Function for a  Class of Processes \\
       Involving  Non Local  Self - Replication}

  \author{ T. Padmanabhan}
 \affil{Inter-University Centre for Astronomy and Astrophysics, Post Bag 4, Ganeshkhind, Pune 411007, India}
 \email{nabhan@iucaa.ernet.in}
 
  \begin{abstract}
  A  large class of evolutionary processes can be modeled by  a rule
that involves self-replication  of some physical quantity with a non local
rescaling.  
We show that a class of such models  are exactly solvable ---
in the discrete as well as continuum limit --- and can represent several
physical situations as varied  from the formation of galaxies in some 
cosmological models to growth of bacterial cultures.
 This class of models, in general,  has no steady state solution and evolve unstably as $t \to \infty$ for generic initial conditions. They can however exhibit  (unstable)   power law correlation
function  in the continuum limit, for an intermediate range of times and length scales.
  \end{abstract}  
  

 \section{Introduction}
 
 Observations show that the two point correlation function of the  galaxies 
 is an approximate power law over a range of scales. 
 This result, known for decades now, defies a simple ``first-principle" explanation because of
 the complexity of the physical processes involved.  In the conventional big-bang cosmology
 the dominant contribution to the energy density of the universe is in the form of nonbaryonic dark matter
 and the visible galaxies (made of baryons) form due to complex processes of cooling and fragmentation
within  the dark matter halos. 
 It is possible to model non linear
 gravitational clustering of {\it dark matter halos} by a scaling ansatz (see eg. \citet{tp96}) and show that 
  a power law correlation
  is expected (in a spatial and temporal interval) for scale invariant initial
 conditions. This result arises  from the fact that Newtonian gravitational 
 dynamics in a $\Omega =1$ universe does not have any preferred scale 
 which is of interest to cosmology. But to translate this result to baryonic structures is not easy 
 and at present
 the only explanation for the power law correlation of galaxies arises from numerical simulations.
 
 There are, however, alternate (and less  popular)  models for galaxy formation in 
 quasi-steady state cosmology (QSSC) in which
 the process is addressed without invoking gravitational instability
 explicitly \citep{nayeri99}. It has been claimed, based on numerical simulations, that 
   these models also lead to power law correlation function for galaxy distribution. 
   This result is of intrinsic interest  even to conventional cosmologists --- who may
   not accept the  alternate model  ---  because it provides a scenario to probe the relationship
   between (i) the conventional scenario of galaxy formation via gravitational instability (accepted by
   most of the working cosmologists) and (ii) the power law correlation function for the galaxies. If a
   completely different  model can lead to power law correlation function, then it is clear that
   observations of galaxy correlation function, by itself, cannot be used as a discriminator between the
   theories.
   
   Since the ``rule'' used for creating galaxies in QSSC
   is simple and explicit, it is indeed possible to study this process
   completely analytically. We provide such an analysis in this paper and show that
   the model is intrinsically unstable. However, the galaxies produced by 
   this rule {\it will} exhibit a power law correlation function for a range of 
   intermediate time scales.

 The rest of the paper is arranged as follows. In the next section, we provide an analytic description of the
 model for galaxy formation in the QSSC, which was used in \citep{nayeri99}.  This is done in terms of
 discrete time steps to parallel the previous work. In section 3 we discuss the continuum limit of this model
 that lends itself to a completely tractable analysis.  In section 4 we describe a very general 
 class of processes which obey the same equations as described here and a broader class of applications.
 The last section gives the conclusions.
 
 \section{A model for structure formation - discrete version}
 
  A non standard model for galaxy formation  in QSSC is built along the following lines:
   Consider a set of points in the 3-dimensional space which represents the location of  galaxies
   at a given instant of time. We now 
generate a set of  new galaxies near each one of the galaxies (``non-local self-replication'')
by a physical process which should quite minimally involve some extra negative energy
field. 
The key ingredient of the nonstandard cosmological models  is the existence of some ``creation field''
so that galaxies could generate new galaxies
nearby
  (see \citet{nayeri99}).  Let
the probability
 for any given new galaxy to be located at a distance ${\bf l}$ from an
old galaxy be $W({\bf l}) d^3{\bf l}$.  
This process will increase the number of galaxies as the universe evolves. In QSSC, a balance
between creation and quasi-steady state structure is maintained by rescaling the size of 
the universe in each cycle. Mathematically, this is taken into account by rescaling 
 each of the 3 dimensions
by a factor $\bar \mu >1$ thereby increasing the volume of available space.
   (If we create
{\it one}  new galaxy near an old one with a probability $W({\bf l})$, 
then we will be doubling the number of galaxies during the self replication.
The rescaling should be such that the volume doubles giving 
$\bar \mu^3 =2$. For future convenience, we shall keep $\bar\mu$ as 
an arbitrary constant.)
We now  select a subset of  galaxies  in the 
central region such that the total number of galaxies remain the same.
This step renormalizes the process back to the original situation so that the
 process can now be repeated with the new subset of galaxies.
 In the specific case of number of galaxies doubling per cycle, we should take
 the central region containing half the volume.
   It is obvious that such a process of galaxy formation will lead to correlations between
   the galaxies since new galaxies are created close to the old ones with a probability
   $W({\bf l})$. The key question is how the correlation function scales with the 
   distance scale. We will now provide a mathematical analysis of this problem.

The
evolutionary rule  described above can be stated mathematically in the   form:
\begin{eqnarray}
 Q(n +1, {\bf x}) &=& {1 \over (1 + \lambda ) \bar\mu^3} \Big[ Q (n, {\bf x}/\bar\mu) \nonumber \\ 
 &&\qquad +\quad  \lambda \int Q (n, {{\bf x} \over \bar \mu} - {\bf l}) W ({\bf l})d {\bf l} \Big] \label{threeq}
\end{eqnarray}
where $\lambda, \bar\mu$ are constants with $\lambda >0$ and $\bar\mu>1$;
  $W({\bf l})$ is a probability
function normalized to unity for  integration over all ${\bf l}$. We shall
also assume that $Q$ is normalized in such a way that its integral over
all ${\bf x}$ is unity. This gives us the conditions
\begin{equation}
 \int d {\bf x} \ Q = 1 = \int d {\bf l} \ \ W ({\bf l}).\label{twoint} 
\end{equation}
It is obvious  that equation (\ref{threeq})
 preserves these conditions under evolution because of the explicit normalization
chosen on the right hand side.
 In the context of the quasi steady state cosmological model, $Q$ will be the ratio between the 
number density of galaxies and the mean density. The normalization in (\ref{twoint}) preserves the 
quasi steady state condition between different cycles under the simultaneous action of 
matter creation and expansion.

  While equation (\ref{threeq}) was motivated from a particular model for galaxy formation,
  the rest of the analysis only uses this equation and is independent of the assumptions
  which go into it. In section 4 we will comment on a wider class of phenomena  which
  could be modeled by such an equation.

Let us now consider the solutions to equation (\ref{threeq}).  Linearity in $Q$ suggests switching to the Fourier space variables $f(n, {\bf k})$ with 
\begin{equation}
 f (n, {\bf k}) = \int d {\bf x}\  Q (n, {\bf x}) e^{i {\bf k} . {\bf x}} ; \quad f (n, 0) = 1. 
\end{equation}
Equation  (\ref{threeq}) now reduces to a simple form
\begin{equation}
 f(n +1, {\bf k}) = {f(n, \bar\mu {\bf k}) \over (1 + \lambda)} \left[ 1 + \lambda W({\bf k}) \right] \label{threek}
 \end{equation}
where we have denoted by the same symbol $W({\bf k}) $ the Fourier transform of the  probability function. Given
the form of this probability function, this equation iteratively
determines the evolution of $f$.

While the equation is fairly  simple in structure, it is not 
easy to find its general solution. Note that the conservation condition 
 (\ref{twoint}) --- which demands $f(n,0) = 1$ for all $n$ --- is satisfied in this case 
because $W({\bf k}=0)=1$ for a normalized probability distribution. This also shows that  solutions of the form
$f(n,{\bf k})=A({\bf k}) \exp(\alpha n)$, with some constant $\alpha$, are unacceptable
for $\alpha \ne 0$ because they will violate the normalization condition. 
Further, since we have the freedom to choose the initial
condition $f(0,{\bf k}) = f_{\rm in} ({\bf k})$, any general solution to 
equation (\ref{threek}) must  contain one arbitrary function; it is 
difficult to obtain such  a {\it general} solution.

A special class of solutions to equation (\ref{threek})  will correspond to a steady
state such that $f(n+1,{\bf k})=f(n,{\bf k}) = f_{\rm s}({\bf k})$.
This function satisfies the equation
\begin{equation}
 f_s ({\bf k}) = f_s ({\bf k} \bar \mu) \left[ {1 + \lambda W ({\bf k}) \over 1 + \lambda }\right].\label{twof}
 \end{equation}
 Two trivial solutions to this equation correspond to $f_s({\bf k})=(0,\infty)$
and, in fact, we shall see later that the most generic initial conditions will  
drive the system to either of these two limits by our process. The only nontrivial solution
which exists can be found by the iteration in the form
\begin{eqnarray}
 f_s({\bf k} \bar\mu )& = & {f_s ({\bf k}) (1 + \lambda) \over 1 + \lambda W ({\bf k}) } = {f_s ( {{\bf k} \over \bar\mu} )(1 + \lambda)^2 \over \left[ 1 + \lambda W ({\bf k} ) \right] \left[ 1 + \lambda W ({{\bf k} \over \bar\mu}) \right]}\nonumber \\
&=&f_s \left( {{\bf k} \over \bar \mu^N}\right) \prod^N _{n=0} {( 1 + \lambda) \over   1 + \lambda W ({\bf k}/ \bar\mu^n ) }.
 \end{eqnarray}
Taking the logarithms, followed by the limit $N\to\infty$, and using $f_s(0)=1$
we get the result
\begin{equation}
 \ln f_s ({\bf k} \bar \mu) = \sum^\infty _{n=0} \ln \left[{(1 + \lambda) \over 1 + \lambda W ({\bf k} / \bar\mu^n)}\right]. \label{lnsum}
\end{equation}
To study the properties of the solution we note that  $W(k/\bar\mu^n)\approx 1$ 
for all $n>\log _{\bar\mu}(kL)$. This is true for a wide class of probability
distributions with some characteristic scale $L$,
and $W(k)$ usually decreases for $kL\gg 1$. We will also assume that $W(k)$
depends only on the magnitude of ${\bf k}$ because of the statistical
isotropy of the process (though our results are easily generalizable to other cases). It follows that each of the denominators in (\ref{lnsum}) is close to unity for all
$n>\log_{\bar\mu}(kL)$ making all the terms negligibly small for $n>\log_{\bar\mu}(kL)$.
So we need to sum the series in (\ref{lnsum}) only up to $n=n_c\le[\log_{\bar\mu}(kL)]$ with the integer
value, lower than the bound, taken for $n_c$. The result of the sum depends on the
form of $W$ at large $k$. The simplest case corresponds to assuming the $W$
vanishes for $kL\gg 1$ which is exact if we take
\begin{equation}
 W({\bf k}) = \theta (1 - kL) \label{onew}
\end{equation}
where $ \theta (z)=1 $ for $z > 0$ and zero otherwise. Then the asymptotic solution is given by
\begin{equation}
f_s (\mu k) = (1 + \lambda)^{(n_c +1)} = (kL)^{\gamma}; \qquad  \gamma =
 {\ln (1 + \lambda) \over \ln \bar\mu} 
\label{twolns} 
\end{equation}
(The equivalence of the two forms follows from simple algebra and definition
of $n_c$.) This is a power law solution with the index determined essentially
 by the two parameters of the problem.

  Before proceeding further, let us consider the effect of  the normalization condition
  (\ref{twoint}) more closely. 
  This is of some interest because one might feel that such a normalization is unnecessarily 
  restrictive and a wider class of phenomena can be modeled by relaxing this condition.
  It turns out, however, that our qualitative considerations are not affected by this constraint.
  Relaxing this normalisation can
   be done most conveniently by replacing the factor $(1+\lambda)$
  in the denominator of the right hand side
  of (\ref{threeq}) by $(1+\lambda_1)$ where $\lambda_1$ is a constant different from $\lambda$.
  This will change equation (\ref{threek}) to the form:
  \begin{equation}
 f(n +1, {\bf k}) = {f(n, \bar\mu {\bf k}) \over (1 + \lambda_1)} \left[ 1 + \lambda W({\bf k}) \right].
  \label{newthreek}
 \end{equation}
 Setting ${\bf k}=0$ and using $W(0)=1$, we get 
 \begin{equation}
   f(n+1,0) = f(n,0) \left({1+\lambda\over 1+\lambda_1}\right) 
   = f(1,0) \left({1+\lambda\over 1+\lambda_1}\right)^n. 
 \end{equation}
   Since the integral over all space of $Q(n,{\bf x})$ is just $f(n,0)$ we find that
   \begin{equation}
   \int d{\bf x} \, Q(n+1,{\bf x}) = \left({1+\lambda\over 1+\lambda_1}\right)^n  \int d{\bf x} \, Q(1,{\bf x}). 
   \label{forq}
   \end{equation}
   This shows how the total number of galaxies changes with time
   if $\lambda\neq \lambda_1$. For example,  a cosmological model in which the mean densities of 
   galaxies decreases with time can be modeled with $\lambda_1 > \lambda$ and interpreting
   $Q$ as the number density of galaxies. In this case, equation (\ref{newthreek}) admits
   solutions of the form $f(n,{\bf k})=A({\bf k}) \exp(\alpha n)$ with
   \begin{equation}
   A({\bf k}) = A (\bar\mu {\bf k})  \left({1+\lambda W({\bf k})\over 1+\lambda_1}\right) e^{-\alpha}.
   \end{equation}
   This equation has the same form as (\ref{twof}) with the factor $(1+\lambda)$ replaced
   by $(1+\lambda_1) e^\alpha$ so that the solution is
   \begin{equation}
 A({\bf k} \bar\mu ) =A \left( {{\bf k} \over \bar \mu^N}\right) \prod^N _{n=0} {( 1 + \lambda_1) e^\alpha\over   1 + \lambda W ({\bf k}/ \bar\mu^n ) }.
 \label{fora}
 \end{equation}
Taking the logarithms, followed by the limit $N\to\infty$, 
we get the result
\begin{equation}
 \ln A ({\bf k} \bar \mu) = \sum^\infty _{n=0} \ln \left[{(1 + \lambda_1) e^\alpha\over 1 + \lambda W ({\bf k} / \bar\mu^n)}\right] +\ln A(0).
 \label{newlnsum}
\end{equation}
   The convergence of the product in the right hand side (\ref{fora})  is now more tricky. If $W ({\bf k} / \bar\mu^n)$
   becomes close to unity for sufficiently large $n$, then we will pick up a factor $p=(1+\lambda_1)
   (1+\lambda)^{-1}e^\alpha$ in each of the terms. Unless this factor is unity, the product in
   (\ref{fora}) will either diverge or vanish. Thus we get the condition $p=1$ for convergence
   thereby making equation (\ref{fora}) identical to (\ref{twof}).
   Thus, when the normalization condition in (\ref{twoint}) is relaxed, we again get
   the same $k-$dependence as in (\ref{twolns}) with an extra time ($n$) dependence
   of the form 
   \begin{equation}
   \exp(\alpha n)=\left({1+\lambda \over 1+\lambda_1}\right)^n
   \label{extrafactor}
   \end{equation}
   which takes into
   account the condition  (\ref{forq}). This clearly shows that nothing changes qualitatively
   as far as the spatial dependence is concerned by relaxing (\ref{twoint}).
   
    We thus have three possible steady state solutions $f_s=[0,\infty, (kL)^\gamma]$
{\it none} of which incorporates {\it arbitrary} initial conditions. Obviously if the system is started at any of these solutions, it will stay in it without any evolution. 
The question arises as to whether any of them acts as a fixed point for the
system evolving from a {\it nontrivial}  initial condition or even random Poisson initial conditions for which $| f_{in} (k)|^2 =1 $. While this question is difficult to
analyze in the discrete model, it 
can be answered using a
 much more detailed description of the system in the continuum limit.

 \section{A model for structure formation - continuum limit}

In the continuum limit we need to study the system at 
two infinitesimally separated moments in time $t$ and $t+\Delta t$ and obtain
a partial differential equation for the evolution of $Q(t,x)$ or --- equivalently
--- for $f(t,k)$. We will also need to change the parameters $\lambda$ and
$\bar\mu$ to $\lambda\Delta t$ and $(1+\mu\Delta t)$ respectively for consistency; this has the effect of making the {\it rate} of creation and {\it rate} of stretching finite, as it should.
Equation (\ref{twof})  now becomes
\begin{equation}
f(t + \Delta t, {\bf k}) = {f(t,{\bf k} (1+ \mu \Delta t) ) \over 1 + \lambda} \left[ 1 + \lambda W(k) \right].
\label{newsix}
\end{equation}
 We have assumed that $W$ depends only on $|{\bf k}|$ making $f$ also depend only on $|{\bf k}|$. Expanding the equation retaining up to linear terms in $\Delta t$, and using the
result
\begin{equation}
 k^{\alpha} {\partial f(k) \over \partial k^{\alpha}} = k {\partial f \over \partial k}
 \end{equation}
 we get the final partial differential equation satisfied by $f$ to be
\begin{equation}
 {\partial f \over \partial t} - \mu k {\partial f \over \partial k} = - \lambda f (1 - W). 
 \label{partialone}
\end{equation}
The general solution to this equation is straightforward to obtain; we find that
\begin{equation}
 \ln f(t, k) = G(ke^{\mu t}) + {\lambda \over \mu} \int^k_0 {dq \over q} \left[ 1- W(q) \right] 
 \end{equation}
 where $G$ is an arbitrary function of its argument with the condition
$G(0)=0$. 
[This condition incorporates our normalization condition (\ref{twoint}). This condition 
can be relaxed in exactly the same manner as in the discrete case; however, as we shall see 
later, 
our conclusions do not change.]
This function - in turn -  can be expressed in terms of the initial 
condition for the problem $f(t=0,k)\equiv f_{in}(k)$, which is assumed to be known.
Doing this we can write the solution in the form
\begin{equation}
 f(t, k) = f_{\rm in} (ke^{\mu t}) \exp \left[ - {\lambda \over \mu} \int^{ke^{\mu t}} _k {dq \over q} \left[ 1 - W(q)\right] \right]. \label{dqq}
 \end{equation}
 We shall now study the properties of the solution.

Let us first consider the simple case in which $W$ is given by equation (\ref{onew}). 
In this case the solution is found to be
\begin{equation}
 f(t,k) =\cases{f_{in}(ke^{\mu t})  &($ k \leq L^{-1}e^{-\mu t})$\cr
f_{in}(ke^{\mu t}) e^{-\lambda t} (kL)^{-\lambda /\mu} & ($ L^{-1}e^{-\mu t} \leq k \leq L^{-1})$ \cr
f_{in} (ke^{\mu t}) e^{-\lambda t} &($  L^{-1} \leq k)$ \cr}
\end{equation}
 At any finite $t$, there is a range of $k$ values for which the power spectrum
(which is proportional to $|f|^2$) is a power law with the index $(-2\lambda/\mu)$).
But note that as $t\to\infty$ the solution decays exponentially at all scales
for a wide class of initial conditions (including Poisson distribution with  $f_{in} = 1$). More generally, if $f_{in}(k)\propto k^\beta$, then
the solution in the three ranges go as
\begin{equation}
f(t,k) \propto\cases{ k^{\beta} \exp (\beta \mu t) \cr
k^{(\beta - {\lambda \over \mu})} \exp \left[ \mu (\beta - {\lambda \over \mu})t \right] \cr
k^{\beta} \exp \left[ \mu (\beta - {\lambda \over \mu} )t \right].\cr}
\end{equation}
 For generic values of the parameters $(\beta, \lambda, \mu)$  {\it all} the solutions tend to either
zero or infinity at late times. The only exception is if we choose the initial
spectrum with $\beta=(\lambda/\mu)$. Then, for $kL<1$, we get a
pure power law $k^{(\lambda/\mu)}$. This is precisely the solution we found in the
discrete case [see equation (\ref{twolns})], since in the continuum limit, we can set
\begin{equation}
\gamma \equiv {\ln (1 + \lambda) \over \ln \bar\mu }\Biggl|_{\rm discrete} \rightarrow {\ln (1+ \lambda \Delta t) \over \ln (1 + \mu \Delta t) } = {\lambda \over \mu}. 
 \end{equation}
 It must be emphasised that only a very special choice can lead to a nontrivial steady state solution. Even with this special choice we get a solution
which has {\it no}  time dependence at all. It is, however, possible to have a
power law solution that is  valid for a large range of $k$ at any finite $t$ but with amplitude decreasing exponentially.
 
Equation (\ref{dqq}) can be explicitly integrated for several cases of $W$. A particularly
simple case is the one with
\begin{equation}
W ({\bf l}) \propto \exp (-l/L); \qquad W({\bf k}) = {1 \over 1+ k^2L^2}. 
\end{equation}
In this case the solution is given by
\begin{equation}
 f(t,k) = f_{\rm in} (ke^{\mu t}) \left[ {1 + k^2L^2 \over 1 + k^2L^2e^{2 \mu t}}\right]^{{\lambda \over 2 \mu}}.
\end{equation}
As $\mu t \to \infty$, the solution goes to
\begin{equation}
f(t \rightarrow \infty, k) \simeq f_{in} ( \infty ) \left( 1 + k^2L^2 \right)^{\lambda /2 \mu}(kL)^{-\lambda /\mu} e^{-\lambda t} 
\end{equation}
 which is again in conformity with the results obtained above, thereby showing that
they were  not an artifact of the sharp cutoff assumed in equation (\ref{onew}).

  Finally, we briefly mention the results in the continuum limit when the normalisation
  condition (\ref{twoint}) is relaxed. In this case, the factor $(1+\lambda)$ in the denominator
  on the right hand side of equation (\ref{newsix}) will be replaced by a new constant
  $(1+\lambda_1)$. This changes equation (\ref{partialone}) and the solution (\ref{dqq})
  to 
\begin{equation}
 {\partial f \over \partial t} - \mu k {\partial f \over \partial k} = - \lambda f \left({\lambda_1\over \lambda}
  - W\right), 
\end{equation}
 \begin{equation}
 f(t, k) = f_{\rm in} (ke^{\mu t}) \exp \left[ - {\lambda \over \mu} \int^{ke^{\mu t}} _k {dq \over q} 
 \left[  {\lambda_1\over \lambda}- W(q)\right] \right]. \label{newdqq} 
 \end{equation}
  Writing the factor inside the integral as
  \begin{equation}
  \left({\lambda_1\over \lambda}
  - W\right) = (1-W) + {(\lambda_1 - \lambda)\over \lambda},
  \end{equation}
  it is trivial to see that the solution in this case is the same as the solution
  with $\lambda_1=\lambda$ multiplied by the factor $\exp [-(\lambda_1 - \lambda)t]$.
  That is, 
  \begin{equation}
  f(t,k) |_{\lambda_1\ne\lambda} = f(t,k) |_{\lambda_1 =\lambda}\exp [-(\lambda_1 - \lambda)t].
  \label{newfff}
  \end{equation}
  This is exactly what one would expect based on our results in the discrete limit 
  in which we found that the solution gets multiplied by a factor in (\ref{extrafactor}).
  In the continuum limit, $\lambda_1 $ and $\lambda$ gets replaced by
  $\lambda_1 \Delta t$ and $\lambda \Delta t$ and $n$ becomes
  $t/\Delta t$. In the limit of $\Delta t \to 0$ the extra factor  in   (\ref{extrafactor})  becomes
  \begin{eqnarray}
  \lim_{\Delta t\to 0} \left( {1+\lambda\Delta t\over 1+\lambda_1 \Delta t}\right)^{t\over \Delta t} &=& 
  \lim_{\Delta t\to 0} \exp \left[ \left( {t\over \Delta t}\right)
  \ln   \left( {1+\lambda\Delta t\over 1+\lambda_1 \Delta t}\right)\right] \nonumber \\
  &=& \exp [-(\lambda_1 - \lambda) t]
  \end{eqnarray}  
  which is precisely the extra factor in (\ref{newfff}). Thus, even in the continuum limit, our 
  qualitative conclusions  do not change when the condition
  (\ref{twoint}) is relaxed.
  
   \section{A generic class of physical processes with non-local self-replication}
   
   While the discussion above was modeled and motivated by  galaxy formation in
   QSSC, the process described here has a much broader range of applicability.  The fact that 
   our results were obtained for a model that does
   not involve gravity explicitly (for example, neither Newtonian 
   gravitational constant nor the fact that gravitational force varies as 
   a power law is used in the ``rule"  mentioned above), suggests
   that results of the above kind could be quite general.
  
   In fact, there exists several natural phenomena  that
   are described by  power law correlations (see eg. \citet{mandel83}). More often than not, such
   a correlation function 
   seem to arise in a manner that does not depend critically
   on the details of the underlying dynamical model. It would be 
   interesting to see whether one can provide a mathematical model with a 
   minimal set of assumptions which can reproduce the power law
   correlation. One such minimal set of  assumptions can be extracted from
   the above analysis and we can show that
   models based on these assumptions will have a generic behaviour.
   
   Consider a dynamical process in which some physical quantity $Q(t,{\bf x})$
evolves in time in a manner which depends on its value non locally. Such an evolution can be treated in the discrete version in terms of a rule which 
allows one to compute $Q(n+1,{\bf x})$ in terms of $Q(n,{\bf x})$ where
$n$ represents the discretised version of time with, say, $t=n\tau$ and $\tau$
representing a convenient time interval. 
For example, the amount of bacteria in a culture, trees in an orchard, buildings in a city or the number of galaxies in some region of the universe
could be studied by such prescription. To be more specific, we shall consider processes of the 
following kind: We start with
  a set of points in the D-dimensional space which represents the location of bacteria or galaxies or
 trees, say. We now 
generate a set of  new points near each one of the points (``non-local self-replication''). 
(The bacteria creating new bacteria nearby or trees generating new trees nearby or even cities leading to the formation of new cities nearby seem reasonable.) Let
the probability
 for any given new point to be located at a distance ${\bf l}$ from an
old point is $W({\bf l}) d^D{\bf l}$.  Next, we rescale each of the $D$ dimensions
by a factor $\bar \mu >1$ thereby increasing the volume of available space. (This is useful in the case of growing bacterial culture or a jungle of trees or cities in order to avoid boundary effects which will limit the process.
In the cosmological model mentioned above, this is a natural consequence of the expansion of the universe.)
Finally we select a subset of particles in the 
central region such that the total number of particles remains the same.
This step renormalizes the process back to the original situation so that the
 process can now be repeated with the new subset of points. 
This
evolutionary rule can be stated mathematically in the   form:
\begin{eqnarray}
 Q(n +1, {\bf x}) &=& {1 \over (1 + \lambda ) \bar\mu^D} \Big[ Q (n, {\bf x}/\bar\mu) \nonumber \\ 
 &&\qquad +\quad  \lambda \int Q (n, {{\bf x} \over \bar \mu} - {\bf l}) W ({\bf l})d {\bf l} \Big] \label{geneq}
\end{eqnarray}
where $\lambda, \bar\mu$ are constants with $\lambda >0$ and $\bar\mu>1$; $D$ is the dimension
of the space in which the vector ${\bf x}$ lives, and $W({\bf l})$ is a probability
function normalized to unity for  integration over all ${\bf l}$. 
This is same as equation (\ref{threeq}) with $\bar \mu^3$ replaced by $\bar \mu^D$.
We shall
also assume that $Q$ is normalized in such a way that its integral over
all ${\bf x}$ is unity. As discussed above, this can be easily relaxed if required.

 It must be stressed that we can consider equation  (\ref{geneq}) as
the basic postulate of this analysis rather than any physical model described in the last
paragraph (involving bacteria, trees, cities, galaxies ....). In particular: (i) we need not restrict to any specific form of $W({\bf l})$ or a choice
for dimension $D$; (ii) it is also not necessary to identify the vector
${\bf x}$ with the position vector in real space. The equation (\ref{threeq})  can
also describe quite effectively the power transfer in Fourier space when the 
vector ${\bf x}$ is actually identified with a Fourier space vector.  There are  phenomena, like fluid turbulence, in which our analysis
can be applied by using the power spectrum as the basic variable in ${\bf k}$
space. The essential postulate will then be that power at nearby wave numbers is generated
with a given probability. This could provide a  tool for attacking a
wide class of nonlinear phenomena.

\section{Conclusions}

  Two point correlations functions which exhibit power law behaviour are very prevalent
  in nature. The analysis here suggests that two ingredients -- which we have called 
  (i) non local self replication and (ii) rescaling -- can lead to such correlation functions
  fairly generically. 
  
  The first of these two ingredients (non local self replication) allows
  similar entities to be created at nearby locations in some space. This could be
  as varied as bacterium creating new bacterium nearby by cell division, trees
  creating trees nearby by seeding or galaxies creating fresh galaxies because
  of the existence of a creation field. If we take the space to be the Fourier 
  domain, then transfer of power from a given wave number $k$ to nearby
  wave numbers will also constitute such a process. This variety shows
  that the ingredient (i) is fairly generic and natural. 
  
  An immediate consequence
  of self replication as defined here is the development of correlations.
  This is because we have tacitly assumed that the members of the second generation  are preferentially (in the probabilistic sense)
  produced near the original parents. This will naturally lead to the second 
  generation to be correlated with the first, spatially. If the process takes place
  in the Fourier domain, then the correlation will arise in the space of wave numbers
  and the consequent relationship in real space will be more complicated; but even in this
  case correlation will definitely be established. (Of course, in principle one can also
  introduce negative correlation in the model by choosing the probability distribution
  to be an increasing function of distance near the origin.) 
  
  In the context of conventional 
  galaxy formation due to gravitational instability, the development of correlation 
  is more indirect and dynamical while the process described here is direct and kinematical.
  Ultimately, the kinematics of the model, encoded in the probability distribution function 
  $W({\bf l}) $, need to be connected to an underlying model (say, the creation field in
  the case of QSSC) in order to provide the dynamical basis. While this is the 
  basic paradigm in physics, it must be stressed that it suffers from two well
  known difficulties: (a) If the physical process is sufficiently complicated (fluid
  turbulence, galaxy formation ...), it just may not be possible to provide
  such a dynamical underpinning. (b) Modeling different processes separately
  could lead one to miss the existence of a very general description for widely
  different classes of phenomena. Much of the initial attraction
  for fractals in the description of natural phenomena originated from its promise
  to provide such a unifying perspective. The mathematical formalism
  developed here should be viewed against such a backdrop.
  
  The second ingredient that we have used (rescaling) essentially serves to renormalise the 
  scales and is, in fact, very similar to ideas used in the theory of renormalization group.
  Once self replication takes place, the system has become denser on the whole and in 
  order to concentrate on the intrinsic correlation, it is necessary to renormalise the system
  back to the original state. In all the examples which we have  it is always possible to link this rescaling with a tangible physical process. 
  
  The combination of 
  these two ingredients very nicely and naturally leads to a system with higher level of 
  correlation at each time step.  In fact, at least in the case of galaxy formation, the combined
  effect of these two phenomena is very similar to actual gravitational attraction between the 
  particles.
  
  The key result of the analysis is that processes with these two ingredients are inherently unstable
  in the sense that the correlation function either grows without bound or decays to zero (with the 
  system becoming more and more dilute) as $t\to \infty$. Once again such an instability is reminiscent
  of similar phenomena seen in self gravitating systems though we have not used any gravitational
  dynamics. There is one special initial condition which leads to a static solution but as we discussed
  in the text, this is rather too  special. The interest in these two systems lies in the intermediate time
  scale during which it could exhibit a power law correlation function very generically. Once again
  the situation is similar to galaxy formation due to gravitational instability in which the observed 
  power law correlation function will exist {\it only} during a limited temporal and spatial window
  in the numerical simulations. Even in the case of fractals, it is generally known that one has
  to introduce cut-off in spatial and temporal scales in order to maintain  power law correlation 
  functions. 
  
  In conclusion, it is interesting how a fairly simple mathematical model could lead to 
  an approximate description of a wide class of phenomena. Only further investigations for 
  specific contexts will determine whether such a unified treatment is of 
  some value or whether these ideas are destined to remain as mathematical curiosities.

I thank J.V. Narlikar for comments on an earlier version of the manuscript.
I appreciate the  detailed and constructive comments from an anonymous referee which significantly
improved the presentation of the ideas.

 \end{document}